\documentclass[prl,aps,twocolumn,superscriptaddress]{revtex4-1}
\usepackage[dvips]{graphicx}
\usepackage[british]{babel}
\usepackage{amsmath,amssymb,amsthm,dsfont,hyperref,ctable,url,braket,mathtools}
\usepackage[percent]{overpic}
\input{qcircuit}

\newcommand{\Tr}{\operatorname{Tr}}
\newcommand{\conv}{\operatorname{conv}}
\newcommand{\rank}{\operatorname{rank}}

\newcommand{\pp}{\operatorname{Pos}}

\newcommand{\N}[1]{\left|\!\left|{#1}\right|\!\right|}

\newtheorem{prop}{Proposition}

\begin{document}

\title{Device-Independent Tests of Quantum States}

\author{Michele \surname{Dall'Arno}}

\email{cqtmda@nus.edu.sg}

\affiliation{Centre for  Quantum Technologies, National
  University of  Singapore, 3 Science Drive  2, 117543,
  Singapore}

\date{\today}

\begin{abstract}
  We construct a  correspondence between quantum states
  and the observable input-output correlations they are
  compatible  with.  The  problem is  framed as  a game
  involving  an experimenter,  claiming to  be able  to
  prepare some  family of  states, and  a theoretician,
  whose  aim  is  to  falsify such  a  claim  based  on
  observed correlations only. For any such a claim, the
  optimal  strategy consists  of providing:  i) to  the
  experimenter,   {\em  all}   the  measurements   that
  generate extremal input-output  correlations, and ii)
  to the theoretician,  the {\em full} characterization
  of  such  correlations.  Comparing  the  correlations
  observed   in  i)   with  those   predicted  by   ii)
  corresponds   to  device-independently   testing  the
  states.  While  no assumption is made  about the {\em
    actual}  states  and  measurements, we  derive  the
  optimal strategy  in {\em  closed-form} for  the case
  when the {\em claim} consists of qubit states and the
  performed measurements are tests, and as applications
  we specify  our results  to the case  of any  pair of
  pure states and to the  case of pure states uniformly
  distributed on the Bloch equatorial plane.
\end{abstract}

\maketitle

Quantum systems are most generally described by quantum
states, abstract  vectors in a mathematical  space with
the property of not  being perfectly distinguishable --
a property  called {\em superposition} of  pure states.
However,  all an  observer can  ultimately observe  are
just   correlations  among   perfectly  distinguishable
events in usual space and time.  Hence, how can quantum
states be  inferred?  Here, we answer  this question by
constructing  a correspondence  between quantum  states
and the  observable input-output correlations  they are
compatible with.

The  problem  is  most   generally  framed  as  a  game
involving  an  experimenter,  claiming to  be  able  to
prepare  $m$  quantum  states  $\{ \rho_x  \}$  and  to
measure them, and a skeptical theoretician whose aim is
to falsify such a  claim based on observed correlations
only.  At  each  run   of  the  experiment,  first  the
experimenter prepares state $\rho_x$ upon input of $x$,
and then  performs measurement  $\{ \pi_{y|w}  \}$ upon
input  of  $w$.   Finally,  the  theoretician  collects
outcome    $y$,    thus   reconstructing    correlation
$\{ p_{y|x,w} \}$.  The setup is as follows:
\begin{align*}
  p_{y|x, w} := \Tr[\rho_x \pi_{y|w} ] \quad = \quad
  \begin{aligned}
    \Qcircuit @C=8pt @R=4pt {
      \push{x \;}  & \prepareC{\rho_x} \cw & \qw & \multimeasureD{1}{\pi_{y|w}} & \cw & \push{\; y} \\
      & & \push{w \;} & \cghost{\pi_{y|w}} }
  \end{aligned} \;.
\end{align*}

Let  us  denote  with  $S_n(   \rho_x  )$  the  set  of
correlations generated by states $\{ \rho_x \}$ for any
$n$-outcomes measurement $\{ \pi_y \}$, that is
\begin{align*}
  S_n(\rho_x)  :=  \left\{  p  \; \Big|  \;  p_{y|x}  =
    \Tr[\rho_x \pi_y] \right\}
\end{align*}
(we  take $y  \in [0,  n-2]$ since  for $y  = n-1$  one
simply has $p_{n-1|x} = 1 - \sum_{y=0}^{n-2} p_{y|x}$).
On  the theoretician's  side,  the  problem amounts  to
fully characterizing $S_n(\rho_x)$,  for any $\{ \rho_x
\}$,  in  order  to  check  if  $\{  p_{y|x,w}  \}  \in
S_n(\rho)$, for  any $w$.  On the  experimenter's side,
the  problem  amounts   to  choosing  measurements  $\{
\pi_{y|w} \}$ generating  all the extremal correlations
of  $S_n(\rho_x)$  (of  course,  the  validity  of  the
conclusion itself will be  independent of $\{ \pi_{y|w}
\}$).   Therefore, $w$  represents  a  direction to  be
probed  in  the  space  of  correlations  in  order  to
reconstruct  $S_n(\rho_x)$.   Since,  as  shown  later,
$S_n(\rho_x)$ is  strictly convex, $w$ is  a continuous
parameter.

\begin{figure}[hbt]
  \includegraphics[width=.5\columnwidth]{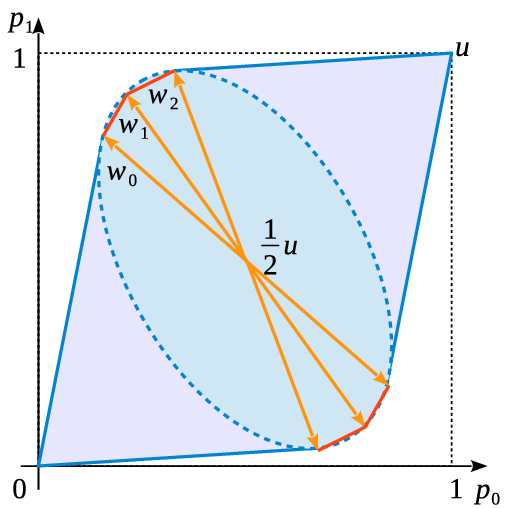}
  \caption{Our    results    admit   a    bidimensional
    geometrical representation for the  case of $m = 2$
    states  $\{ \rho_x\}$.   For any  direction $w$
    (yellow vectors)  in the space of  correlations, we
    provide the experimenter  with the measurements $\{
    \pi_{y|w} \}$ generating a correlation that lies on
    the  boundary  of  $S_2(\rho_x)$  if  the  measured
    states are $\{ \rho_x \}$.  To check if this is the
    case,  we  provide  the theoretician  with  a  full
    closed-form   characterization  of   $S_2(\rho_x)$,
    which  turns out  to be  given by  the convex  hull
    (blue area) of $0$  and $u$ and the ellipse
    (blue       dashed       line)       given       by
    Eqs.~\eqref{eq:preview}.}
  \label{fig:ellipse}
\end{figure}

Here, we  provide a  full closed-form solution  of this
problem  for the  case when  the claim  $\{ \rho_x  \}$
consists of qubit states --  notice that this is a {\em
  restriction} on  the claim to be  tested, rather than
an {\em  assumption} on  the actual  states --  and the
performed measurements are  tests, that is measurements
with  $n =  2$ outcomes.   In particular,  for any  $\{
\rho_x \}$,  we explicitly derive: i)  the measurements
$\{  \pi_{y|w}  \}$  generating a  correlation  at  the
boundary  of  $S_2(\rho_x)$  for {\em  any  arbitrarily
  given  direction}   $w$;  and   ii)  the   {\em  full
  closed-form  characterization} of  $S_2(\rho_x)$.  It
turns  out that  $S_2(\rho_x)$ is  given by  the convex
hull of  the two isolated  points $0$ and  $u$ (vectors
with  null  and  unit entries,  respectively)  and  the
ellipsoid given by the system:
\begin{align}
  \label{eq:preview}
  \begin{cases}
    (\openone - Q^{-1}Q) (p - \frac12 u) = 0,\\
    (p -  \frac12 u)^T Q^{-1} (p  - \frac12
    u) \le 1,
  \end{cases}
\end{align}
where $Q_{x_0,x_1} = \frac12 \Tr[\rho_{x_0} \rho_{x_1}]
-   \frac14$.   This   situation   is  represented   in
Fig.~\ref{fig:ellipse}.
As applications,  we explicitly discuss the  case where
$m =  2$ and $\{  \rho_x \}$  are pure states,  and the
case  where  $\{  \rho_x  \}$ are  distributed  on  the
$m$-vertices  of   a  regular  polygon  on   the  Bloch
equatorial plane.

Our  results share  analogies  with  previous works  on
device-independent       testing       of       quantum
dimension~\cite{GBHA10,   HGMBAT12,  ABCB12,   DPGA12}.
Notice  however  that therein  the  aim  is to  test  a
specific  scalar  property  of states  $\{  \rho_x  \}$
rather then  their most  general operatorial  form, and
the set of correlations  is probed along an arbitrarily
chosen direction rather than being fully reconstructed.
Moreover, the present author has recently addressed the
very  related problems  of device-independent  tests of
quantum  channels~\cite{DBB16,  BD18,  APCSCBDS18}  and
measurements~\cite{DBBV16, DBBT18}.

{\em Experimental observations.}  ---  We will make use
of   standard  definitions   and  results   in  Quantum
Information Theory~\cite{NC00}.   Any quantum  state is
represented  by  a density  matrix  $\rho$,  that is  a
unit-trace   positive   semi-definite  operator.    Any
quantum    measurement    is     represented    by    a
positive-operator  valued  measure  (POVM), that  is  a
collection  $\{  \pi_y  \}$ of  positive  semi-definite
operators  such that  $\sum_y \pi_y  = \openone$.   The
conditional probability $p_{y|x}$  of outcome $y$ given
input state $\rho_x$ is given by the Born rule, that is
$p_{y|x} = \Tr[\rho_x \pi_y]$.

The experimenter  claims to  be able to  prepare states
$\{ \rho_x \}$  and to measure them.  Their  task is to
support such claims by  generating all the correlations
at the boundary of $S_n(\rho_x)$.  To this aim, for any
direction  $w$  in  the   space  of  correlations,  the
experimenter must  measure the  POVM $\{  \pi_{y|w} \}$
that generates  the correlation $p_{y|x}  := \Tr[\rho_x
\pi_{y|w}]$ that  maximizes $p^T w$.  In  this section,
we derive any such a POVM  for any given $\{ \rho_x \}$
and $w$.

Formally, $\{ \pi_{y|w} \}$ is given by the solution of
the following optimization problem:
\begin{align}
  \label{eq:threshold0}
  W(\rho_x, w)  := \max_{\substack{\{ \pi_y \ge  0\} \\
      \sum_y  \pi_y  =  \openone}} \sum_{x  =  0}^{m-1}
  \sum_{y=0}^{n-2} w_{x,y} \Tr[\rho_x \pi_y].
\end{align}

In  the following,  we make  the restriction  $n =  2$,
hence $p$ and $w$ are  column vectors with $m$ entries.
Therefore, the maximum  in Eq.~\eqref{eq:threshold0} is
attained when  $\pi_0$ is the projector  on $\pp(\sum_x
w_x  \rho_x)$,   where  $\pp(\cdot)$   denotes  the
positive part  of operator $(\cdot)$, and  in this case
one has
\begin{align}
  \label{eq:threshold1}
  W(\rho_x, w) = \Tr\left[  \pp \left(\sum_x w_x \rho_x
    \right) \right].
\end{align}

Hence,   our  first   result  provides   a  closed-form
characterization   of  the   POVM  $\{   \pi_{y|w}  \}$
achieving  the  correlation  $p$  at  the  boundary  of
$S_2(\rho_x)$  that maximizes  $p^T w$,  for any  given
family $\{ \rho_x\}$ of states and direction $w$.

\begin{prop}
  \label{prop:povms}
  For any family $\{ \rho_x \}$ of states and direction
  $w$  in  the  space  of correlations,  the  POVM  $\{
  \pi_{y|w} \}$ generating  the correlation $p_{y|x} :=
  \Tr[\rho_x   \pi_{y|w}]$    on   the    boundary   of
  $S_2(\rho_x)$  that maximizes  $p^T w$  is such  that
  $\pi_{0,w}$ is  the projector  on $\pp  \left( \sum_x
    w_x  \rho_x \right)$  and $\pi_{1|w}  = \openone  -
  \pi_{0|w}$.
\end{prop}

Proposition~\ref{prop:povms} restricts the set of POVMs
$\{  \pi_{y|w}\}$ that  need to  be measured.   Indeed,
whenever  $\pp \left(  \sum_x w_x  \rho_x \right)$  has
rank zero or two,  the corresponding correlation $p$ is
trivial (i.e.  $p = 0$  or $p = u$, respectively), thus
direction $w$ does not need to be probed.

{\em  Theoretical predictions.}   --- The  theoretician
does  not  believe  any  of  the  claims  made  by  the
experimenter   about   the   experimental   setup,   in
particular  about the  set of  POVMs $\{  \pi_{y|w}\}$.
Their  task is  to test  such claims  by comparing  the
observed correlations with $S_2(\rho_x)$.  To this aim,
in  this   section  we   provide  a   full  closed-form
characterization of $S_2(\rho_x)$ under the restriction
that $\{ \rho_x \}$ are qubit states.

The   set  $S_2(\rho_x)$   is   recovered  by   further
optimizing    $W(\rho_x,   w)$,    as   given    by
Eq.~\eqref{eq:threshold1}, over  any direction $w$,
that is:
\begin{align}
  \label{eq:program0}
  S_2 \left( \rho_x \right) = \left\{ p \; \Big| \; p =
  \max_w  \left( p^T  w -  W \left(  \rho_x, w  \right)
  \right) \le 0 \right\}.
\end{align}

Upon fixing  a computational basis, $\{  \rho_x \}$ can
be decomposed  in terms of Pauli  matrices $\{ \sigma_k
\}$ as follows
\begin{align*}
  \rho_x  =  \frac12  \openone +  \sum_{k=1}^3  S_{x,k}
  \sigma_k,
\end{align*}
where  $S_{x,k} :=  \frac12  \Tr[\rho_x \sigma_k]$.  Of
course, our result will be independent of the choice of
computational basis.

It is then a simple computation to find that
\begin{align*}
  W(\rho_x, w)  = \max \left(0, \;  \N{w}_1, \;
    \frac12 \N{w}_1 + \N{S^Tw}_2 \right),
\end{align*} 
where  $\N{\cdot}_p$  denotes  the $p$-norm  of  vector
$(\cdot)$.   The   maximum  is  achieved  by   $0$  and
$\N{w}_1$ if $\{ \pi_{y|w} \}$ is trivial ($\pi_{0|w} =
0$ and  $\pi_{0|w} =  \openone$, respectively),  and by
$\frac12 \N{w}_1 + \N{S^Tw}_2$  if $\{ \pi_{y|w} \}$ is
rank-one projective.  If $\{  \pi_{y|x} \}$ is trivial,
the  optimization  problem  in  Eq.~\eqref{eq:program0}
becomes
\begin{align*}
  \begin{cases}
    \max_{w} p^T w \le 0, & \quad \textrm{if } \pi_0 = 0,\\
    \max_{w} (p - \frac12 u)^T w \le 0,
    & \quad \textrm{if } \pi_0 = \openone
  \end{cases}    
\end{align*}
which, as expected, are satisfied  if and only if $p
= 0$ and $p = u$, respectively.

If however  $\{ \pi_{y|w}  \}$ is  rank-one projective,
the  optimization  problem  in  Eq.~\eqref{eq:program0}
becomes
\begin{align}
  \label{eq:program1}
  \max_{w} \left[ (p - \frac12 u)^T w -
    \N{S^T w}_2 \right] \le 0.
\end{align}
This optimization problem is  formally equal to that in
Eq.~(5)  of Ref.~\cite{DBBV16},  where  the problem  of
device-independent  tests of  quantum measurements  was
addressed.    Notice  however   that  the   operational
interpretation   and,  accordingly,   the  mathematical
representation  of  the  symbols  are  different.   For
example,  in  Ref.~\cite{DBBV16}   $p$  represents  the
probability distribution of the outcomes of a POVM, and
thus $\sum_y  p_y = 1$,  while here $p$  represents the
vector of probabilities of outcome $\pi_0$ given states
$\rho_x$, and thus there is no linear constraint on the
sum of its elements. Analogous differences hold for $u$
($t$ in  Ref.~\cite{DBBV16}) and $S$.  The consequences
of    these   differences    on    the   solution    of
Eq.~\eqref{eq:program1} will be discussed at the end of
this section.

Since Eq.~\eqref{eq:program1} is  left invariant by the
transformation $w  \to w  = |(p -  \frac12 u)^T
w|^{-1} w$  (we recall  that $w$ only  represents a
direction in  the space of correlations),  without loss
of  generality one  can  take $(p-\frac12  u)^T
w = 0, \pm 1$.  When $(p-\frac12 u)^T w
= 0, - 1$, the inequality in Eq.~\eqref{eq:program1} is
of course satisfied,  thus let $(p-\frac12 u)^T
w = 1$.  Equation~\eqref{eq:program1} becomes
\begin{align}
  \label{eq:program2}
  \min_{\substack{w   \\  (p-\frac12   u)^T
      w = 1}} \N{S^T w}_2^2 \ge 1,
\end{align}
that  is, a  linearly-constrained quadratic-programming
problem.

Let us denote with $Q$  the real symmetric matrix $Q :=
SS^T$.   Upon   denoting    with   $(\cdot)^{-1}$   the
Moore-Penrose   pseudoinverse~\cite{AG03}   of   matrix
$(\cdot)$,  one has  that $\openone  - Q^{-1}Q$  is the
orthogonal projector on the kernel of $Q$.

Let us first show that a necessary condition for $p \in
S_2(\rho_x)$ is that  $p - \frac12 u$  is orthogonal to
the kernel of $Q$.  Indeed,  suppose by absurd that $(p
- \frac12  u)^T (\openone  - Q^{-1}Q) (p  - \frac12
u) > 0$. By setting
\begin{align*}
  w = \frac{(\openone - Q^{-1}Q) \left( p - \frac12
      u  \right) }{  \left( p  - \frac12  u
    \right)^T  (\openone  -  Q^{-1}Q)  \left(  p  -
      \frac12 u \right)}.
\end{align*}
one  immediately has  that  the constraint  $(p-\frac12
u)^T  w =  1$ is  verified, and  that $\N{S^T  w}_2^2 =
0$.  Therefore,  by Eq.\eqref{eq:program2}  $p  \not\in
S(\rho_x)$.

Let then $p  - \frac12 u$ belong to the  kernel of $Q$.
In this  case, we can  take without loss  of generality
$w$ in  Eq.~\eqref{eq:program2} to have support  on the
kernel  of $Q$.   Then,  it  is known~\cite{BV04}  that
Eq.~\eqref{eq:program2} is solved by
\begin{align}
  \label{eq:solution}
  \begin{cases}
    Q w  = - \lambda (p - \frac12 u),\\
    (p - \frac12 u)^T w = 1,
  \end{cases}
\end{align}
where $\lambda$ is a Lagrange multiplier. The system in
Eq.~\eqref{eq:solution} is solved by~\cite{AG03}
\begin{align}
  \label{eq:solution2}
  \begin{cases}
    w = - \lambda Q^{-1} (p - \frac12 u) + (\openone - Q^{-1}Q) v,\\
    (p  - \frac12  u)^T \left[  (\openone -  Q^{-1}Q) v
      -\lambda Q^{-1} (p - \frac12 u) \right] = 1.
  \end{cases}
\end{align}

If $(p -  \frac12 u)^T Q^{-1} (p - \frac12  u) > 0$, by
taking $v = 0$,
\begin{align*}
  \lambda  =  \left[ \left(  p  -  \frac12 u  \right)^T
    Q^{-1} \left( p - \frac12 u \right) \right]^{-1},
\end{align*}
and $w = \lambda Q^{-1} (p  - \frac12 u)$, one has that
the system in  Eq.~\eqref{eq:solution2} is verified, as
well  as the  constraint $(p  -  \frac12 u)^T  w =  1$.
Hence, $w$ is the  solution of the optimization problem
in   Eq.~\eqref{eq:program2},  and   one  has   $\N{S^T
  w}_2^2 = \lambda$, that is $p \in S_2(\rho_x)$ if
and only if  $(p - \frac12 u)^T Q^{-1} (p  - \frac12 u)
\le  1$. If  instead $(p  -  \frac12 u)^T  Q^{-1} (p  -
\frac12 u) = 0$, one has $p = \frac12 u$, that is again
$p \in S_2(\rho_x)$.

Hence,  the solution  of  the  optimization problem  in
Eq.~\eqref{eq:program2} is given by
\begin{align}
  \label{eq:ellipse}
  \begin{cases}
    (\openone - Q^{-1}Q) (p - \frac12 u) = 0,\\
    (p -  \frac12 u)^T Q^{-1} (p  - \frac12
    u) \le 1.
  \end{cases}
\end{align}
Finally, by explicit computation it immediately follows
that   $Q$  is   given   by   $Q_{x_0,x_1}  =   \frac12
\Tr[\rho_{x_0} \rho_{x_1}] - \frac14$, thus as expected
the system in Eq.~\eqref{eq:ellipse} does not depend on
the choice of computational basis.

Then,   our  second   main  result   provides  a   full
closed-form characterization  of the  set $S_2(\rho_x)$
of  correlations compatible  with  any arbitrary  given
qubit family $\{ \rho_x\}$ of states.

\begin{prop}
  \label{prop:characterization}
  The set $S_2(\rho_x)$ of  correlations generated by a
  given family $\{  \rho_x \}$ of qubit  states and any
  test $\{ \pi_y \}$ is given by
  \begin{align*}
    S_2(\rho_x) = \conv  \left\{ 0, u, \textrm{
        Eq.~\eqref{eq:ellipse}} \right\},
  \end{align*}
  where    $Q_{x_0,x_1}   =    \frac12   \Tr[\rho_{x_0}
  \rho_{x_1}] - \frac14$.
\end{prop}

Let   us  provide   a  geometrical   interpretation  of
Proposition~\ref{prop:characterization}.  The system of
equalities    in   Eq.~\eqref{eq:ellipse}    represents
$\rank(\openone -  Q^{-1}Q)$ linear  constraints, while
the inequality  represents an  $m$-dimensional cylinder
with $(\rank Q)$-dimensional hyper-ellipsoidal section.
Thus,   Eq.~\eqref{eq:ellipse}  represents   a  $(\rank
Q)$-dimensional   hyper-ellipsoid    embedded   in   an
$m$-dimensional space.  Since $\rank  Q \le 3$, we have
that  Eq.~\eqref{eq:ellipse}  respresents  a  (possibly
degenerate)  ellipsoid. Notice  as  a comparison  that,
while  in  this  case $S_2(\rho_x)$  includes  the  two
isolated correlations $0$  and $u$, in the  case of the
device-independent        tests       of        quantum
measurements~\cite{DBBV16} no isolated correlations are
included.

{\em   Comparison.}  ---   Finally,   we  discuss   the
comparison of  the set of correlations  observed by the
experimenter according  to Proposition~\ref{prop:povms}
and the set $S_2(\rho_x)$ predicted by the theoretician
according  to  Proposition~\ref{prop:characterization}.
Notice first  that the inclusion  relation $S_2(\rho_x)
\supseteq  S_2(\rho'_x)$  induces  a  partial  ordering
among families of quantum states $\{ \rho_x \}$ and $\{
\rho'_x \}$, that  is $\{ \rho_x \}  \succeq \{ \rho'_x
\} \Leftrightarrow S_2(\rho_x) \supseteq S_2(\rho'_x)$.
Of  course, if  the  experimenter  produces {\em  some}
correlation not in $S_2(\rho_x)$, the theoretician must
conclude that  the prepared states $\{  \rho'_x \}$ are
such that
\begin{align}
  \label{eq:weaktest}
  \{ \rho'_x \} \not\prec \{ \rho_x\}.
\end{align}

However,  if the  experimenter produces  {\em all}  the
extremal   correlations   of  $S_2(\rho_x)$   (as   per
Proposition~\ref{prop:povms}),  the  theoretician  must
conclude that  the prepared states $\{  \rho'_x \}$ are
such that
\begin{align}
  \label{eq:strongtest}
  \{ \rho'_x \} \succeq \{ \rho_x \}.
\end{align}
Since    the    ordering    $\succeq$    is    partial,
Eq.~\eqref{eq:strongtest}   is   of   course   strictly
stronger   than    Eq.~\eqref{eq:weaktest},   that   is
Eq.~\eqref{eq:strongtest}                       implies
Eq.~\eqref{eq:weaktest}  but the  vice-versa is  false.
Informally,   Eq.~\eqref{eq:strongtest}    allows   the
theoretician to  lower bound the ``ability''  to create
input-output correlations of the states prepared by the
experimenter.

An  even stronger  result  can be  achieved  when $m  =
2$.              In              this              case
Proposition~\ref{prop:characterization}   provides  for
the first  time the  full closed-form  quantum relative
Lorenz curve  for any  pair $\{  \rho_0, \rho_1  \}$ of
qubit state, as  illustrated by Fig.~\ref{fig:ellipse}.
Quantum  relative  Lorenz  curves  have  been  recently
introduced  by  Buscemi  and  Gour~\cite{BG17}  in  the
context  of   quantum  relative  majorization.    As  a
consequence of  a result  therein, in  turn based  on a
previous  result  by Alberti  and  Uhlmann~\cite{AU80},
under  the  additional  assumption  that  the  prepared
states  $\{  \rho'_0,  \rho'_1 \}$  are  qubit  states,
Eq.~\eqref{eq:strongtest}  implies the  existence of  a
quantum    channel    $\mathcal{C}$,    that    is    a
completely-positive  trace-preserving linear  map, such
that
\begin{align}
  \label{eq:alberti}
  \mathcal{C}(\rho'_x) = \rho_x, \qquad x = 1,2.
\end{align}
Therefore, Eq.~\eqref{eq:alberti} means that the states
$\{ \rho'_x  \}$ prepared by the  experimenter are less
noisy than the claimed states $\{ \rho_x \}$.  However,
it is known~\cite{Mat14} that this implication fails if
the assumption that the  prepared states $\{ \rho'_x\}$
are qubit states is relaxed.

{\em Applications.}  --- As an application  of the case
$m = 2$, we consider any  pair of pure states $\rho_x =
\ket{\psi_x} \!  \!  \bra{\psi_x}$, that can be written
without loss of generality as
\begin{align*}
  \ket{\psi_0}   =  \ket{0},   \qquad  \ket{\psi_1}   =
  \cos\frac\alpha2 \ket{0} + \sin\frac\alpha2 \ket{1}.
\end{align*}
Since           $|\braket{\psi_0|\psi_1}|^2           =
\cos^2\frac\alpha2$,  matrix  $Q_{x_0,x_1}  :=  \frac12
|\braket{\psi_{x_0}|\psi_{x_1}}|^2 -  \frac14$ is given
by $Q = [(1+\cos\alpha) v_+v_+^\dagger + (1-\cos\alpha)
v_-v_-^\dagger]/4$,  where $v_\pm  = 1/\sqrt{2}(1,  \pm
1)^T$.   If  $\alpha  \neq   0,  \pi$,  the  system  in
Eq.~\eqref{eq:ellipse} becomes
\begin{align*}
  \frac1{1+\cos\alpha}          (p_0+p_1-1)^2         +
  \frac1{1-\cos\alpha} (p_0-p_1)^2 \le \frac12.
\end{align*}
If  $\alpha   =  0$   or  $\alpha   =  \pi$,   that  is
$\ket{\psi_0}         =        \ket{\psi_1}$         or
$\braket{\psi_0|\psi_1} =  0$ respectively,  the system
in Eq.~\eqref{eq:ellipse} trivially becomes $p_0 = p_1$
or $p_0 = 1 - p_1$, respectively.

As an application  of the general case  we consider $m$
pure states $\rho_x  = \ket{\phi_x} \!\!  \bra{\phi_x}$
uniformly  distributed in  the Bloch  equatorial plane,
that can be written without loss of generality as
\begin{align*}
  \ket{\phi_x} =  \cos \frac{\pi  x}{m} \ket{0}  + \sin
  \frac{\pi x}{m} \ket{1}.
\end{align*} 
Since   $|\braket{\phi_{x_0}|\phi_{x_1}}|^2  =   \cos^2
\frac{\pi   (x_0-x_1)}{m}$,   matrix  $Q_{x_0,x_1}   :=
\frac12  |\braket{\phi_{x_0}|\phi_{x_1}}|^2 -  \frac14$
is circulant, that is  $Q_{x_0+k, x_1+k} = Q_{x_0,x_1}$
for  any  $x_0$,  $x_1$,  and $k$.   Therefore,  it  is
lengthy but not difficult  to show that its eigenvalues
are given by
\begin{align*}
  \lambda_j   &   =   \frac14   \sum_{k=0}^{m-1}   \cos
  \frac{2\pi k}{m} \exp\frac{2 \pi i j (m-k)}{m} \\ & =
  \frac{\left(  e^{2   \pi  i  j}  -   1\right)  \left(
      e^{\frac{2 \pi i j}m} + e^{\frac{2 \pi i (j+2)}m}
      -   2    e^{\frac{2   \pi   i}m}    \right)}   {8
    \left(e^{\frac{2\pi  i}m} -  e^{\frac{2 \pi  i j}m}
    \right)   \left(  e^{\frac{2   \pi  i   (j+1)}m}  -
      1\right)}.
\end{align*}
Hence, one  has that $\lambda_1 =  \lambda_{m-1} = m/8$
and  $\lambda_j =  0$ otherwise,  and two  eigenvectors
$v_\pm$ corresponding to non-null eigenvalues are given
by $v_\pm$ where $(v_\pm)_k := \frac1{\sqrt{m}}
\exp   \left(  \pm   \frac{2   \pi   i  k}m   \right)$.
Accordingly,  one  has  that   $Q  =  \frac{m}8  \left(
  v_+v_+^\dagger    +    v_-v_-^\dagger
\right)$,  and  the  system  in  Eq.~\eqref{eq:ellipse}
becomes
\begin{align*}
  \begin{cases}
    (\openone - v_+v_+^\dagger - v_-v_-^\dagger) p = 0,\\
    \N{v_+^\dagger p}_2^2 \le \frac{16}m.
  \end{cases}
\end{align*}
For  instance,  consider  the   case  of  two  mutually
unbiased  bases~\cite{KR05} (MUBs),  obtained for  $m =
4$.    MUBs  have   applications  e.g.    in  classical
communications   over  quantum   channels~\cite{Dal14},
quantum   cryptography~\cite{BB84},   and  locking   of
classical         information        in         quantum
states~\cite{DHLST04}.  One has  that $v_\pm  = (1,
\pm  i,  -1,  \mp  i)^T$,  from  which  the  system  in
Eq.~\eqref{eq:ellipse} becomes
\begin{align*}
  \begin{cases}
    p_0 +  p_2 = p_1  + p_3  = 1,\\
    \N{p}_2^2 \le \frac32.
  \end{cases}
\end{align*}

{\em Conclusion.}   --- In this work  we have addressed
the  problem of  constructing a  correspondence between
any given family  $\{ \rho_x \}$ of  $m$ quantum states
and  the set  $S_n(\rho_x)$ of  observable correlations
they  can generate  for  any POVM  $\{  \pi_y \}$.  The
problem  has  been  framed   as  a  game  involving  an
experimenter,  claiming  to  be able  to  prepare  some
family $\{  \rho_x \}$  of states, and  a theoretician,
willing to  trust observed correlations only.   For any
such  a  claim $\{  \rho_x  \}$,  the optimal  strategy
consists  of providing:  i)  to  the experimenter,  the
measurement   $\{  \pi_{y|w}   \}$  that   generates  a
correlation on  the boundary of $S_n(\rho_x)$  for {\em
  any}   given   direction   $w$,  and   ii)   to   the
theoretician,  the   {\em  full}   characterization  of
$S_n(\rho_x)$.  Comparing the  correlations observed in
i)  with   those  predicted   by  ii)   corresponds  to
device-independently  testing  the  states.   While  no
assumption has been made  about the {\em actual} states
and measurements, we have  derived the optimal strategy
in {\em closed-form} for the  case when the {\em claim}
consists of qubit states and the performed measurements
are tests, that is measurements  with $n = 2$ outcomes,
and  discussed the  geometrical  interpretation of  our
results.   As  applications,   we  have  specified  our
results to the  case of any pair of pure  states and to
the case  of pure  states uniformly distributed  on the
Bloch equatorial plane.

Natural  open problems  include  relaxing  some of  the
restrictions  we  considered, e.g.   considering  POVMs
with  arbitrary  number  of   outcomes  and  states  in
arbitrary dimension.  Furthermore, the characterization
of  the set  $S_n(\rho_x)$  of correlations  compatible
with an arbitrary dimensional  family $\{ \rho_x \}$ of
$m =  2$ states might  prove to be  the key to  solve a
well-known       longstanding       conjecture       by
Shor~\cite{Sho02}, based on numerical work by Fuchs and
Peres: whether the accessible information of any binary
ensemble is  attained by  a Von Neumann  POVM. Finally,
the  full closed-form  characterization of  the quantum
relative  Lorenz curve  for  qubit  states provided  by
Proposition~\ref{prop:characterization} naturally leads
to      applications      in      quantum      resource
theories~\cite{DB17},  within   the  general  framework
provided by the quantum Blackwell theorem~\cite{Bus12}.

We conclude by noticing that our results are remarkably
suitable  for  experimental  implementation.   For  any
family of  qubit states that an  experimenter claims to
be  able to  prepare, our  framework only  requires Von
Neumann  measurements  to  be  performed  in  order  to
experimentally reconstruct  the entire boundary  of the
set  of compatible  correlations.

{\em  Acknowledgements.}  ---  M.~D.  is  grateful to  Sarah
Brandsen  and Francesco  Buscemi  for valuable  discussions,
comments, and  suggestions.  This  research is  supported by
the National  Research Fund  and the Ministry  of Education,
Singapore,   under  the   Research  Centres   of  Excellence
programme.


\begin{thebibliography}{}
\bibitem{GBHA10} R.  Gallego,  N.  Brunner, C.  Hadley,
  and  A.   Ac\'in,  {\em Device-Independent  Tests  of
    Classical  and  Quantum  Dimensions},  Phys.   Rev.
  Lett.  {\bf 105}, 230501 (2010).
\bibitem{HGMBAT12}  M.    Hendrych,  R.    Gallego,  M.
  Mi\v{c}uda,   N.    Brunner,   A.  Ac\'in,   and   J.
  P.  Torres,  {\em   Experimental  estimation  of  the
    dimension of classical and quantum systems}, Nature
  Phys.  {\bf 8}, 588-591 (2012).
\bibitem{ABCB12}  H.    Ahrens,  P.    Badzi\c{a}g,  A.
  Cabello,  and   M.   Bourennane,   {\em  Experimental
    Device-independent Tests  of Classical  and Quantum
    Dimensions}, Nature Physics {\bf 8}, 592 (2012).
\bibitem{DPGA12}   M.   Dall'Arno,   E.   Passaro,   R.
  Gallego, and  A.  Ac\'in,  {\em Robustness  of device
    independent  dimension  witnesses}, Phys.  Rev.   A
  {\bf 86}, 042312 (2012).
\bibitem{DBB16}   M.   Dall'Arno,   S.   Brandsen,   F.
  Buscemi,  {\em  Device-independent tests  of  quantum
    channels},  Proc. R.  Soc.  A  {\bf 473},  20160721
  (2017).
\bibitem{BD18}  F.   Buscemi  and M.   Dall'Arno,  {\em
  Data-driven Inference of Physical Devices: Theory and
  Implementation}, arXiv:1805.01159.
\bibitem{APCSCBDS18}   I.    Agresti,   D.    Poderini,
  G. Carvacho,  L.  Sarra, R.  Chaves,  F.  Buscemi, M.
  Dall'Arno,  and  F.    Sciarrino,  {\em  Experimental
    semi--device--independent    tests    of    quantum
    channels},    Quantum   Science    and   Technology
  \textbf{4}, 035004 (2019).
\bibitem{DBBV16}  M.    Dall'Arno,  S.    Brandsen,  F.
  Buscemi,  and  V.   Vedral,  {\em  Device-independent
    tests        of       quantum        measurements},
  Phys. Rev. Lett. {\bf 118}, 250501 (2017).
\bibitem{DBBT18}  M. Dall'Arno,  F. Buscemi,  A. Bisio,
  and   A.   Tosini,    {\em   Data-Driven   Inference,
    Reconstruction,  and Observational  Completeness of
    Quantum Devices}, arXiv:1812.08470.
\bibitem{NC00} M.  A.  Nielsen and I.  L.  Chuang, {\em
    Quantum   computation   and  quantum   information}
  (Cambridge university press, 2010).
\bibitem{BV04}  S.~P.   Boyd,  L.   Vandenberghe,  {\em
    Convex  Optimization} (Cambridge  University Press,
  2004).
\bibitem{AG03}  B.-I.   Adi, T.~N.~E.   Greville,  {\em
    Generalized Inverses} (Springer-Verlag, 2003).
\bibitem{KR05} A.  Klappenecker and M.  Roetteler, {\em
    Mutually  unbiased  bases  are  complex  projective
    2-designs},   Proceedings   of    the   2005   IEEE
  International Symposium  on Information  Theory (ISIT
  2005), 1740 (2005).
\bibitem{Dal14}   M.     Dall'Arno,   {\em   Accessible
    Information  and  Informational  Power  of  Quantum
    2-designs}, Phys.  Rev.  A {\bf 90}, 052311 (2014).
\bibitem{BB84} C.   H. Bennett  and G.   Brassard, {\em
    Quantum cryptography:  Public key  distribution and
    coin  tossing}, Proceedings  of IEEE  International
  Conference   on   Computers,   Systems   and   Signal
  Processing {\bf 175}, 8 (1984).
\bibitem{DHLST04} D.  P.  DiVincenzo, M.  Horodecki, D.
  W.  Leung, J.   A.  Smolin, and B.   M.  Terhal, {\em
    Locking classical  information in  quantum states},
  Phys.  Rev.  Lett. {\bf 92}, 067902 (2004).
\bibitem{Sho02} Peter  W.  Shor, {\em On  the Number of
    Elements Needed in a  POVM Attaining the Accessible
    Information}, Quantum Communication, Computing, and
  Measurement {\bf 3}, 107, 2002.
\bibitem{BG17} F.   Buscemi and G.  Gour,  {\em Quantum
    Relative  Lorenz Curves},  Physical  Review A  {\bf
    95}, 012110 (9 January 2017).
\bibitem{AU80} P.  M.  Alberti  and A.  Uhlmann, {\em A
    problem relating to positive  linear maps on matrix
    algebras},  Reports  on Mathematical  Physics  {\bf
    18}, 163 (1980).
\bibitem{Mat14}  K.  Matsumoto,  {\em An  example of  a
    quantum statistical model which cannot be mapped to
    a  less informative  one  by  any trace  preserving
    positive map}, arXiv:1409.5658 (2014).
\bibitem{DB17}   M.    Dall'Arno,   F.    Buscemi,   in
  preparation.
\bibitem{Bus12} F.  Buscemi, {\em Comparison of quantum
    statistical   models:  equivalent   conditions  for
    sufficiency},  Commun.  Math.   Phys.   {\bf  310},
  625-647 (2012).
\end{thebibliography}
\end{document}